\documentclass{article}

\usepackage{amssymb}
\usepackage{amsthm}
\usepackage{amsmath}
\usepackage[active]{srcltx}

\usepackage{graphicx}

\newtheorem{theorem}{Theorem}[section]
\newtheorem{prop}[theorem]{Proposition}

\newtheorem{remark}[theorem]{Remark}

\newcommand{\oz}{{\overline{z}}}
\newcommand{\obeta}{{\overline{\beta}}}

\begin{document}
\title{Elliptic Euler-Poisson-Darboux equation, critical points and integrable systems}
\date{\today}
\author{B.G. Konopelchenko \\
Dipartimento di Matematica e Fisica " Ennio de Giorgi", Universit\`{a} del
Salento \\
INFN, Sezione di Lecce, 73100 Lecce, Italy \\
{konopel@le.infn.it} \\
\mbox{} \\
G. Ortenzi \\
Dipartimento di Matematica Pura ed Applicazioni,\\
Universit\`{a} di Milano Bicocca, 20125 Milano, Italy\\
{giovanni.ortenzi@unimib.it} }
\maketitle

\begin{abstract}
\noindent Structure and properties of families of critical points for classes of functions $W(z,\oz)$ obeying the elliptic  Euler-Poisson-Darboux equation $E(1/2,1/2)$ are studied. General variational and differential equations governing the dependence of critical points in variational (deformation) parameters are found. Explicit examples of the corresponding integrable quasi-linear differential systems and hierarchies are presented There are the extended dispersionless Toda/nonlinear Schr\"{o}dinger hierarchies, the ``inverse'' hierarchy and equations associated with the real-analytic Eisenstein series $E(\beta,\obeta;1/2)$ among them. Specific bi-Hamiltonian structure of these equations is also discussed. 
\end{abstract}
\section{Introduction}
The elliptic Euler-Poisson-Darboux (EPD)  $E(1/2,1/2)$ equation
\begin{equation}
\label{EPD1212}
 (z-\oz) W_{z \oz} -\frac{1}{2}\left(W_z-W_\oz \right)=0
\end{equation}
 where $z=x+iy$ is a complex variable and bar denotes the complex conjugation is of importance in various branches of mathematics and mathematical physics. In the form
\begin{equation}
\label{axialL}
 W_{xx}+W_{yy}+\frac{1}{y}W_y=0
\end{equation}
is the axial three-dimensional Laplace equation, the study of which goes back to Laplace, Stokes \cite{Sto} (see also \cite{Lamb} art. $94$) and Beltrami \cite{Bel}. Equation (\ref{EPD1212}) is the elliptic version of the EPD equation intensively studied in the classical differential geometry (see \cite{Dar}) and in generalizations of the hypergeometric function to the case of two variables \cite{App1}-\cite{Pic}. \par
Its gauge equivalent form 
\begin{equation}
\label{LB14}
 y^2 (u_{xx}+u_{yy})+\frac{1}{4}u=0, \qquad \mathrm{where} \quad u=y^{1/2}W,
\end{equation}
represents the eigenvalue problem for the Laplace-Beltrami operator $y^2(\partial_x^2+\partial_y^2)$ on the Poincar\'e plane for the eigenvalue $-1/4$ corresponding to the beginning of the continuous spectrum (see e.g. \cite{PF,LP}). \par
Equation (\ref{LB14}) is a stationary version of the automorphic wave equation 
\begin{equation}
\label{wave}
 u_{tt}- y^2 (u_{xx}+u_{yy})-\frac{1}{4}u=0
\end{equation}
which is the basic equation for the scattering theory of automorphic function (see e.g. \cite{PF,LP}). Equation (\ref{wave}) is equivalent to the classical wave equation $v_{tt}-v_{xx}-v_{yy}=0$ in $2+1$ dimensional Minkowsky space \cite{LP}. \par
Finally equation (\ref{EPD1212}) (and \ref{axialL}) is the particular instance of the more general EPD $E(k,k)$ equation
\begin{equation}
\label{EPDkk}
 (z-\oz) W_{z \oz} -k\left(W_z-W_\oz \right)=0
\end{equation}
which is the fundamental characterizing equation for the real analytic automorphic forms introduced by Maass \cite{Maa1,Maa2}. Elliptic EPD $E(1/2,1/2)$ equation is also of importance in the representation theory for the $SL(2,\mathbb{R})$ group \cite{GGPS}.\par
Equations equivalent or close to the equation (\ref{EPD1212}) have arisen in various branches of physics (see e.g.\cite{GS}-\cite{DGK}). Recently is was shown that the elliptic EPD $E(1/2,1/2)$ equations quite relevant within the quasiclassical approximation approach for the vortex filament dynamics \cite{KO1,KO2}. In particular it was demonstrated that the solution of the quasi-classical Da Rios system governing the quasi-classical dynamics of vortex filaments describe families of critical points $z_{crit}=\beta$ of a function $W$ solution of the equation (\ref{EPD1212}) with $\beta=-\tau+iK$ where $K$ and $\tau$ and the curvature and the torsion of the filament (\cite{KO1,KO2}). \par
It is hardly necessary to insist on the importance of critical points of functions. For analytic functions they are the points where the map $z'=f(z)$ fails to be conformal while for harmonic functions $u(z,\oz)$ they are the multiple points of level curves and curves of steepest descent (see e.g. \cite{Wal}). In mechanics, critical points of potential are points of equilibrium. In hydrodynamics, they are stagnation points in flow field having $u(z,\oz)$ as velocity potential (see e.g. \cite{Lamb}).
Critical points and degenerate critical points of functions are subject of the catastrophe or singularity theory 
(see e.g. \cite{Thom,AGV}).
The results presented in the papers \cite{KO1,KO2} indicate that the family of critical points of functions obeying EPD equations are described by solutions of integrable hydrodynamic type systems or by elliptic systems of quasilinear PDEs.\par
In the present paper we will study structure and properties of critical points for  families of functions $W(z,\oz)$
 obeying the elliptic EPD $E(1/2,1/2)$ equation (\ref{EPD1212}).  The case $k=1/2$ for the general EPD (\ref{EPDkk}) is an exceptional one and consequently the families of critical points have distinguished properties. \par
First, it is shown that the critical points of a function $W(z,\oz)$ are the double points of the corresponding level curves defined by $W(z,\oz)=W(\beta,\obeta)$. The same is valid for the level curves for the dual function $W^*(z,\oz)$ obeying the EDP equation $E(-1/2,-1/2)$. Moreover these two families of curves form an orthogonal system of coordinates on the plane $z,\oz$. The use of the explicit form of general solution for equation (\ref{EPD1212}), parameterized by two arbitrary functions $\phi(\lambda)$ and $\psi(\lambda)$ allows us to perform rather detailed analysis of the families of critical points for various classes of solutions of equation (\ref{EPD1212}). Explicit  form of variation of $\beta$ and $\obeta$ id found for general variations $\delta \phi(\lambda)$ and $\delta \psi(\lambda)$. It is shown that for general variations it holds
\begin{equation}
 W(\delta \phi, \delta \psi; \beta, \obeta)=\delta F(\beta,\obeta)
\end{equation}
where $F(\beta,\obeta)$ is a certain function.\par
General form of ordinary and partial differential equations governing the dependence of critical points on deformation parameters $x_i,y_i$ is found. It is shown that for the class of solutions of equation (\ref{EPD1212}) of the form $W=\sum_k x_k W_k(z,\oz)$ where $x_k$ are free parameters, the corresponding differential equations for common critical points of a function $W$ and its dual $W^*$ coincide.\par
Particular classes of differential equations and hierarchies associated with specific classes of solutions of  (\ref{EPD1212}) are described explicitly. There are the extended dispersionless Toda/nonlinear Schr\"{o}dinger (dToda/dNLS) hierarchies, ``inverse'' dToda/dNLS hierarchy and equations associated with the functions $\phi$ and $\psi$ of the form of a sum of Dirac delta functions and quasilinear equations related to the 
real-analytic Eisenstein series $E(\beta,\obeta;1/2)$ among them.\par
 Specific bi-Hamiltonian properties of such integrable equations and hierarchies  and application of the general solution of the EPD equation for characterization of general initial data for the quasi-classical Da Rios system are discussed too. \par
General EPD's $E(k,l)$ and associated  structures will be considered in subsequent paper. \par
The paper is organized as follows. In section $2$ some basic known facts about the EPD $E(-1/2,-1/2)$ equation (\ref{EPD1212}), and properties of its solutions are briefly recalled. Properties of critical points for functions $W(z,\oz)$ and its dual function $W^*(z,\oz)$ are discussed in Section 3. Families of critical points and differential equations governing their dependence on deformation parameters are considered in section 4. Concrete examples of integrable differential equations and hierarchies are presented in section 5. Bi-Hamiltonian structure for such equations is discussed in section 6. Application of the obtained results to the vortex filament dynamics in the quasi-classical approximation is considered in section 7.
\section{Properties of the elliptic EPD $\mathbf{E(1/2,1/2)}$ equation}
\label{sec-prop}
First, we will recall some basic properties of equation (\ref{EPD1212}) (see e.g. \cite{Sto}-
\cite{Wei2}) presenting them in a form adapted for our purposes. Equation (\ref{EPD1212}) is invariant under M\"{o}bius transformation
\begin{equation}
 \label{Moeb}
\begin{split}
z \to z'=\frac{az+b}{cz+d}&
W(z,\oz) \to W'(z',\oz')=(cz+d)^{1/2}(c\oz+d)^{1/2}W(z,\oz)
\end{split}
\end{equation}
with real $a,b,c,d$ and $ad-bc=1$ which form the group $SL(2,\mathbb{R})$. In other words the transformation
\begin{equation}
 \label{Appell}
\begin{split}
W(z,\oz) \to W'(z,\oz)=(cz+d)^{-1/2}(c\oz+d)^{-1/2}W\left(\frac{az+b}{cz+d},\frac{a\oz+b}{c\oz+d}\right), 
\qquad a,b,c,d \in \mathbb{R},\quad ac-bd=1
\end{split}
\end{equation}
found by Appell in \cite{App2} converts any solution $W(z,\oz)$ of the equation (\ref{EPD1212}) into a new solution 
$W'(z',\oz')$ of the same equation. In addition, the EPD equation (\ref{EPD1212}) is obviously invariant under the discrete transformation $z \to \oz$, $\oz \to z$  and $W(z,\oz)\to \overline{W(z,\oz)}$.
Equation (\ref{EPD1212}) has the form of the conservation law 
\begin{equation}
\label{claw1212}
 ((z-\oz) W_\oz)_z+((z-\oz) W_z)_\oz=0
\end{equation}
and, consequently  there is a function $W^*$ such that \cite{Bel,Bel2,Wei1}
\begin{equation}
 \label{W*}
W^*_z=(z-\oz) W_z, \qquad W^*_\oz=-(z-\oz) W_\oz.
\end{equation}
The relations (\ref{claw1212}) imply also that 
\begin{equation}
 dW^*=(z-\oz)(W_zdz-W_\oz d\oz)
\end{equation}
and the EPD equation (\ref{EPD1212}) is equivalent to the condition of closeness of differential one-form, i.e. to the equation
\begin{equation}
 d ((z-\oz)(W_zdz-W_\oz d\oz))=0.
\end{equation}
Equation (\ref{EPD1212})
 is the compatibility condition for the system (\ref{W*}) and the function $W^*$ obeys EPD $E(-1/2,-1/2)$ equation 
\cite{Bel}
\begin{equation}
 \label{EPD-12-12}
(z-\oz) W^*_{z \oz} +\frac{1}{2}\left(W^*_z-W^*_\oz \right)=0.
\end{equation}
Equations (\ref{EPD1212}) and (\ref{EPD-12-12}) are dual each other also in the sense that 
\begin{equation}
 \partial_z\partial_\oz L_{-1/2}= L_{-1/2} \partial_z\partial_\oz 
\end{equation}
where $L_k=(z-\oz) \partial_z\partial_\oz -k (\partial_z-\partial_\oz)$.
Equation (\ref{W*})  written in real variables $x$ and $y$ are the generalization of the Cauchy-Riemann equations defining an analytic function and they were one of the origin of the theory of generalized analytic functions \cite{Bel}.
The EPD equation (\ref{EPD1212}) is the Euler-Lagrange equation for the functional 
\begin{equation}
\label{S}
 S=\iint_G (z-\oz) W_z W_\oz\ dz \wedge d\oz.
\end{equation}
This functional is positively-defined for real valued $W$ and for the domain $G$ lying completely in the upper half
plane Im$z>0$.\par
The functional (\ref{S})  represents the positively-defined part of the energy functional 
\begin{equation}
\label{Eu}
 E=\iint_G \left(u_x^2+u_y^2-\frac{u^2}{4y^2} \right) dxdy
\end{equation}
associated with the equations (\ref{LB14}) \cite{PF,LP}. Indeed, passing to the variable $z=x+iy$ and $W=\left(\frac{2i}{z-\oz}\right)^{1/2}u$ one has
\begin{equation}
\label{EW}
 E= \iint_G \left((z-\oz) W_z W_\oz +\frac{i}{2}(W^2)_z-\frac{i}{2}(W^2)_\oz\right)\ dz \wedge d\oz.
\end{equation}
Contribution of the last two terms in the integral (\ref{Eu}) obviously comes from the boundary $\partial G$ while the rest is positively-defined and coincides with (\ref{S}). Variational principle for the EPD equations has been discussed also in \cite{AA}.\par 
Finally we expose some facts about solutions of equation (\ref{EPD1212}). Its general solution is obviously parameterized by two arbitrary functions of a single variable. A convenient form of the general solution is 
\begin{equation}
 \label{W1212}
 W(z,\oz)= \int_{\Gamma_1} d\lambda\ \phi(\lambda) (\lambda-z)^{-1/2}(\lambda-\oz)^{-1/2}+
\int_{\Gamma_2} d\lambda\ \psi(\lambda) (\lambda-z)^{-1/2}(\lambda-\oz)^{-1/2} \ln\left( \frac{z-\oz}{(\lambda-z)(\lambda-\oz)}\right)
\end{equation}
where $\phi$ and $\psi$ are arbitrary functions of the complex variable $\lambda$ and $\gamma_1$, $\Gamma_2$ are two arbitrary contours enclosing the points $z$ and $\oz$. The formula (\ref{W1212}) is the elliptic version of the expression for the general solution of the hyperbolic EPD $E(1/2,1/2)$ equation found by Poisson \cite{Poi} (see \cite{Dar}). \par
Poisson's trick was to start with the general solution of the hyperbolic version of the EPD $E(k,k)$ equation (\ref{EPDkk}). In our case such solution is given by (see e.g. \cite{Dar})
\begin{equation}
 \label{Wkk}
 W(z,\oz)= \int_{\Gamma_1} d\lambda\ \phi_k(\lambda) (\lambda-z)^{-k}(\lambda-\oz)^{-k}+
(z-\oz)^{1-2k} \int_{\Gamma_2} d\lambda\ \psi(\lambda) (\lambda-z)^{k-1}(\lambda-\oz)^{k-1}. 
\end{equation}
In the naive limit $k \to 1/2$ in (\ref{Wkk}) two terms collapse in a single one. Poisson's trick is to choose 
$\phi_{1/2}(\lambda)=\frac{1}{2}\phi(\lambda)+\frac{1}{2\epsilon} \psi(\lambda)$, $\psi_{1/2}(\lambda)=\frac{1}{2}\phi(\lambda)-\frac{1}{2\epsilon} \psi(\lambda)$, $k \to 1/2+\epsilon$ and to take the limit $\epsilon \to 0$ in (\ref{Wkk}). Zero order in $\epsilon$ gives (\ref{W1212}).
A simple way to clarify the Poisson trick is as follows. For small $\epsilon=k-1/2$ and $\Gamma_1=\Gamma_2\equiv \Gamma$ let us choose the functions $\phi_k$ and $\psi_k$ as
\begin{equation}
\label{pp-lin-e}
 \phi_{\frac{1}{2}+\epsilon}(\lambda)=\phi_0(\lambda)+ \epsilon \phi_0(\lambda) \qquad
 \psi_{\frac{1}{2}+\epsilon}(\lambda)=\psi_0(\lambda)+ \epsilon \psi_0(\lambda). 
\end{equation}
For small $\epsilon$ one has
\begin{equation}
 \label{W1212-e}
\begin{split}
 &W_{\frac{1}{2}+\epsilon}(z,\oz)= 
\int_{\Gamma} d\lambda\ (\phi_0(\lambda)+\psi_0(\lambda)) (\lambda-z)^{-1/2}(\lambda-\oz)^{-1/2}
\\
&+\epsilon \int_{\Gamma} d\lambda\  \Big[
\Big(\phi_1(\lambda)+\psi_1(\lambda)-
2 \psi_0(\lambda)\ln(z-\oz)  - (\phi_0-\psi_0 ) \ln((\lambda-z)(\lambda-\oz) )
\Big) (\lambda-z)^{-1/2}(\lambda-\oz)^{-1/2}
\Big]+
o(\epsilon)
\end{split}
\end{equation}
At $\phi_0 \neq -\psi_0$ the zero order in $\epsilon$ term in (\ref{W1212-e}) is dominant and one gets only the first term in (\ref{W1212}), i.e. the naive limit. In order to get the general solution (\ref{W1212}) one should choose $\phi_0 = -\psi_0$. In such a case the first order term in (\ref{W1212-e}) dominates and it reproduces the expression (\ref{W1212}) with $\phi=\phi_1+\psi_1$ and $\psi=2\psi_0$ up to an inessential factor $\epsilon$. Rescaling the functions (\ref{pp-lin-e})by the factor $1/2\epsilon$  and putting $\phi_1=\psi_1$ gives rise the original Poisson trick.\par
Another way to derive the formula (\ref{W1212}) is to use the standard technique for generating solutions of differential equations via symmetry transformation (see e.g.\cite{Wig,Ovs}). First, we observe that equation (\ref{EPD1212}) has two independent solutions $W_0=1$ and $W_0=\ln(z-\oz)$. They represent, in fact, two linearly independent solutions of the equation ${W_0}_{yy}+\frac{1}{y}{W_0}_y=0$ for the zero $x-$Fourier mode of $W(x,y)$
\cite{PF,LP}.\par
Now let us apply the Appell transformations (\ref{Appell}) to those solutions. One gets
\begin{equation}
  \label{AT}
\begin{split}
 {W_0}_1'\equiv AT \cdot {W_0}_1=c^{-1} \left(z+\frac{d}{c}\right)^{-1/2} \left(\oz+\frac{d}{c}\right)^{-1/2} \\
{W_0}_2'\equiv AT \cdot {W_0}_2=c^{-1} \left(z+\frac{d}{c}\right)^{-1/2} \left(\oz+\frac{d}{c}\right)^{-1/2}
\ln\left( \frac{z-\oz}{c^2 \left(z+\frac{d}{c}\right) \left(\oz+\frac{d}{c}\right)}\right)
\end{split}
\end{equation}
where $AT$ indicates the Appell transformation. Denoting $\lambda=-d/c$, multiplying  $ {W_0}_1'$ and $ {W_0}_2'$
by arbitrary coefficients depending on $\lambda$, taking linear superposition of all these solutions with all values of $\lambda$, one gets (\ref{W1212}).\par
The choice $\Gamma_1=\Gamma_2=I$ where $I$ is an interval between $z$ and $\oz$ gives rise to an elliptic version EPD's E($1/2,1/2$) solution reported in book \cite{Dar}. With the parameterization  $\lambda=\cos^2({\alpha})\ z+
\sin^2({\alpha})\ \oz$, $0 \leq \alpha \leq \pi/2$ of $I$ one represents the corresponding solution (\ref{W1212})in the form
\begin{equation}
\label{RDt}
 W(z,\oz)=2i\int_0^{\pi/2}\phi(\cos^2({\alpha}) z+\sin^2({\alpha}) \oz)\ d \alpha+
2i\int_0^{\pi/2}\psi(\cos^2({\alpha}) z+\sin^2({\alpha}) \oz) \ln\left(\frac{(\oz-z)\sin^2(2\alpha)}{4}\right)\ d\alpha.
\end{equation}
In such a form a solution $W(z,\oz)$ of the EPD E($1/2,1/2$) is a sort of Radon transform of functions $\phi$ and $\psi$. Actually if $\psi=0$ the formula (\ref{RDt}) gives the Radon transform of the function $\phi$.\par
The presence of logarithmic terms is the characteristic feature of the general solution (\ref{W1212}). This fact makes the theory of the EPD $E(1/2,1/2)$ equation quite different from the generic case $k\neq 1/2$. \par
Under the $AT$ this general solution transforms according to the formula (\ref{Appell}), namely one has 
\begin{equation}
 \label{W1212'}
\begin{split}
 W'(z,\oz)=&
\int_{{\Gamma'_1}}\frac{d\mu}{\mu+d} \phi\left(\frac{a \mu +b}{c \mu+d}\right) (\mu-z)^{-1/2}(\mu-\oz)^{-1/2}\\ &+
\int_{{\Gamma'_2}}\frac{d\mu}{\mu+d} \psi\left(\frac{a \mu +b}{c \mu+d}\right) (\mu-z)^{-1/2}(\mu-\oz)^{-1/2}
\ln\left( \frac{(z-\oz)(\mu+d)}{(\mu-z)(\mu-\oz)}\right).
\end{split}
\end{equation}
where ${\Gamma'_1}$ and ${\Gamma'_2}$ are preimages of the contours $\Gamma_1$ and $\Gamma_2$ under the M\"obius transformation $\mu \to \lambda = \frac{a \mu+b}{c \mu+d}$. Thus, under $AT$, 
$\Gamma_1 \to {\Gamma_1'}={\Gamma_1}\vert_\mu  \cup {\Gamma_2}\vert_\mu $, 
$\Gamma_2 \to {\Gamma_2'}={\Gamma_2}\vert_\mu  $ and
\begin{equation}
\label{psipsi}
 \psi(\lambda) \to \psi'(\lambda) =\frac{1}{c \lambda+d} \psi\left(\frac{a \lambda +b}{c \lambda+d}\right)
\end{equation}
and $\phi(\lambda) \to \phi'(\lambda)  $ where
\begin{equation}
\label{phiphi}
 \begin{split}
  \phi'(\lambda) \vert_{{\Gamma_1}_mu} =\frac{1}{c \lambda+d} \phi\left(\frac{a \lambda +b}{c \lambda+d}\right) \\
  \phi'(\lambda) \vert_{{\Gamma_2}_mu} =\frac{2 \ln (c \lambda+d)}{c \lambda+d} \psi\left(\frac{a \lambda +b}{c \lambda+d}\right).
 \end{split}
\end{equation}
In particular, for solutions with $\psi \equiv 0$
\begin{equation}
 \phi'(\lambda) =\frac{1}{c \lambda+d} \phi\left(\frac{a \lambda +b}{c \lambda+d}\right).
\end{equation}
Real-valued solutions of the EPD equation (\ref{EPD1212}) invariant under the $AT$ (\ref{Appell}), i.e. those for which
\begin{equation}
\label{aut12}
 W(z,\oz)=(cz+d)^{-1/2}(c\oz+d)^{-1/2}W\left(\frac{az+b}{cz+d},\frac{a\oz+b}{c\oz+d}\right),
\end{equation}
are of particular interest. They are real analytic automorphic forms of weight $1/2$ \cite{Maa1,Maa2}.
\section{Critical points}
Critical points of functions obeying equation (\ref{EPD1212}) have rather specific and remarkable properties.\par
First, let us take a real valued solution $W(z,\oz)$ of the equation (\ref{EPD1212}) and assume that it has $n$ critical points $(\beta,\obeta)$
\begin{equation}
 W_{z}|_{z=\beta,\oz=\obeta}=W_{\oz}|_{z=\obeta,\oz=\obeta}=0.
\end{equation}
Then, we take an arbitrary point $z_0$ on the complex plane and consider the level curve passing through this point. It is defined by the equation 
\begin{equation}
\label{lev-curve-W}
 W(z,\oz)=W_0
\end{equation}
where $W_0=W(z_0,\oz_0)$. This level curve is regular if it does not passes through any critical point $\beta$ of the function $W(z\oz)$. If it passes through the critical point $(\beta_i,\obeta_i)$, i.e. $W(\beta_i,\obeta_i)=W_0$ then this point is obviously the singular point for such level curve.\par
At such point, taking into account that $W_{\beta\obeta}=0$ for $\beta \neq \obeta$ due to equation (\ref{EPD1212}), one has
\begin{equation}
 W_{\beta\beta} (dz)^2+W_{\obeta\obeta} (d\oz)^2=0
\end{equation}
along the level curve near the point $(\beta,\obeta)$. So, for the generic critical point $\beta$ at which $W_{\beta\beta}\neq0$ and $W_{\obeta\obeta}\neq0$, one gets
\begin{equation}
\label{clin-def}
 \left(\frac{d\oz}{dz}\right)^2=-\frac{W_{\beta\beta}}{W_{\obeta\obeta}}.
\end{equation}
Hence, the critical point $\beta,\obeta$ is the double point of the level curve (\ref{lev-curve-W}) with two values of clinant given by
\begin{equation}
 \left(\frac{d\oz}{dz}\right)_1= \left(-\frac{W_{\beta\beta}}{W_{\obeta\obeta}}\right)^{1/2} \qquad
 \left(\frac{d\oz}{dz}\right)_2=-\left(-\frac{W_{\beta\beta}}{W_{\obeta\obeta}}\right)^{1/2}.
\end{equation}
Using the standard formula for the angle between intersecting curves (see e.g. \cite{Dav}),
we conclude that the angle between two arcs of the level curve at the double points is $\pi/2$. \par
For the higher order critical points for which $\frac{\partial^k W}{\partial z^k}\Big{|}_\beta=0$, $\frac{\partial^k W}{\partial \oz^k}\Big{|}_\beta=0$, $k=1,\dots,N$ analog of the relation (\ref{clin-def}) is
 \begin{equation}
\label{clin-def-N}
 \left(\frac{d\oz}{dz}\right)^{N+1}=-\frac{\partial^{N+1} W / \partial \beta^{N+1}}{\partial^{N+1} W / \partial \obeta^{N+1}}.
\end{equation}
Consequently at such critical points the level curve have $N+1$ multiple singular point with the angle $\pi/(N+1)$ between successive arc of the curve.\par
Thus, the level curve (\ref{lev-curve-W}) may have only multiple ordinary singular points of the function $W(z,\oz)$.\par
At the regular points the curve (\ref{lev-curve-W}) can be characterized by the Schwarz function $S(z)$ such that on the curve it holds 
\begin{equation}
 \oz=S(z).
\end{equation}
At the singular points and, hence, at the critical points of the function $W(z,\oz)$, the Schwarz function ceases to exist. \par
Above-mentioned properties of the critical points of real valued solutions of equation (\ref{EPD1212})  are similar, is some respects, to those of harmonic functions. For complex solutions of equation (\ref{EPD1212}) situation is more complicated. Generic critical points of $W(z,\oz)$ are again points of the level curve with the angle $\pi/2$ between two arcs of the curve, while for higher order critical points correspond to more complicated singular points of the curve (\ref{lev-curve-W}).\par
It is  obvious that singular points of the level curves 
\begin{equation}
\label{lev-curve-W*}
 W^*(z,\oz)=W_0^*
\end{equation}
and critical points of the function $W^*$ obeying the dual EPD $E(-1/2,1/2)$ equation (\ref{EPD-12-12}) have the same properties. Moreover, in virtue of (\ref{W*}), a critical point of the function $W(z,\oz)$ is the critical point of the function $W^*(z,\oz)$.\par
Families of the level curves (\ref{lev-curve-W}) and dual level curve (\ref{lev-curve-W*}) form a particular and important family of curves on the plane $(z,\oz)$. First one observes that at each point $(\alpha,\overline{\alpha})$ on the level curve (\ref{lev-curve-W}) there is a dual level curve intersecting the curve (\ref{lev-curve-W}) orthogonally at this point. Indeed, the dual level curve defined by the equation 
\begin{equation}
\label{Walpha}
W^*(z,\oz)=W^*(\alpha,\overline{\alpha})
\end{equation}
by construction intersects the level curve (\ref{lev-curve-W}) at the point  $(\alpha,\overline{\alpha})$. Moreover, in virtues of (\ref{W*}) at the regular point $(\alpha,\overline{\alpha})$ 
\begin{equation}
 \frac{W_\alpha}{W_{\overline{\alpha}}}=
 -\frac{W^*_\alpha}{W^*_{\overline{\alpha}}},
\end{equation}
and, hence, the angle between level curve (\ref{lev-curve-W})  and the dual level curve  (\ref{lev-curve-W*}) at the point $(\alpha,\overline{\alpha})$ is $\pi/2$. So, for each level curve (\ref{lev-curve-W}) there is an infinite family of dual curves (\ref{lev-curve-W*})  intersecting it orthogonally.  Thus, families of regular level curves (\ref{lev-curve-W}) and regular dual curves (\ref{lev-curve-W*}) form an orthogonal family of curves on the plane $z,\oz$ and, hence, provide us with the curvilinear orthogonal system of coordinates. 
 If the point $(\alpha,\overline{\alpha})$ is the generic critical point then it is double point for both level and dual level curves passing through it. in virtue of (\ref{W*}), at this point 
\begin{equation}
  \frac{W_{\alpha\alpha}}{W_{\overline{\alpha}\overline{\alpha}}}=
 -\frac{W^*_{\alpha\alpha}}{W^*_{\overline{\alpha}\overline{\alpha}}},
\end{equation}
and, hence, the clinants of the level and dual level curve s obey the relation 
\begin{equation}
 \left(\frac{d\oz}{dz}\right)^2=-
 {\left(\frac{d\oz}{dz}\right)^*}^2.
\end{equation}
 So, the arcs of the level and dual level curves at such double point are rotated on the angle $\pi/2$ with respect to each other. Therefore the second arc of the level curve touch the first arc of the dual level curve at such point and vice versa. Similar situation takes place for higher order critical points.
\par
Geometric characterization of the critical points of functions $W(z,\oz)$ presented above is supported by their physical meaning: the density of the energy (\ref{EW}) and the functional (\ref{S}) vanishes at the critical points.
\section{Families of critical points and differential equations}
Observations given in the previous section clearly indicate that the analysis of the properties of critical points and families of critical points for functions obeying the EPD equation (\ref{EPD1212}) is of importance.\par
In virtue of formula (\ref{W1212}) critical points $(\beta,\obeta)$ of any function obeying equation (\ref{EPD1212}) are the solutions of equations
\begin{equation}
 \label{crit-eqns}
\begin{split}
 W_{\beta}=&  \frac{1}{2}\int_{\Gamma_1} d\lambda\ \phi(\lambda) (\lambda-\beta)^{-3/2}(\lambda-\obeta)^{-1/2}\\&+
\int_{\Gamma_2} d\lambda\ \psi(\lambda) (\lambda-\beta)^{-3/2}(\lambda-\obeta)^{-1/2} 
\left(\frac{1}{2}\ln\left( \frac{\beta-\obeta}{(\lambda-\beta)(\lambda-\obeta)}\right) +\frac{\lambda-\obeta}{\beta-\obeta}\right)=0,\\
 W_{\obeta}= &\frac{1}{2}\int_{\Gamma_1} d\lambda\ \phi(\lambda) (\lambda-\beta)^{-1/2}(\lambda-\obeta)^{-3/2}\\&+
\int_{\Gamma_2} d\lambda\ \psi(\lambda) (\lambda-\beta)^{-1/2}(\lambda-\obeta)^{-3/2} 
\left(\frac{1}{2}\ln\left(\frac{\beta-\obeta}{(\lambda-\beta)(\lambda-\obeta)}\right) -\frac{\lambda-\beta}{\beta-\obeta}\right)=0.
\end{split}
\end{equation}
Even in the simplest case $\psi\equiv 0$ finding of the solutions for there equations for given $\phi(\lambda)$ is not an easy task. One of the standard way to approach such a problem is to pass from the integral equations (\ref{crit-eqns}) to their consequences in variation. Taking into account that $W_{\beta\obeta}=0$ at $\beta \neq \obeta$, and assuming that $W_{\beta\beta} \neq 0$ and $W_{\obeta\obeta} \neq 0$, one readily gets
\begin{equation}
\label{deltabeta}
 \begin{split}
  \delta \beta =&-\frac{W_{\beta}(\delta \phi,\delta \psi;\beta,\obeta)}{W_{\beta\beta}(\phi,\psi;\beta,\obeta)} =-\frac{1}{W_{\beta\beta}}  \left(\frac{1}{2}\int_{\Gamma_1} d\lambda\ \delta\phi(\lambda) (\lambda-\beta)^{-3/2}(\lambda-\obeta)^{-1/2}\right.\\&+\left.
\int_{\Gamma_2} d\lambda\ \delta\psi(\lambda) (\lambda-\beta)^{-3/2}(\lambda-\obeta)^{-1/2} 
\left(\frac{1}{2}\ln\left( \frac{\beta-\obeta}{(\lambda-\beta)(\lambda-\obeta)}\right) +\frac{\lambda-\obeta}{\beta-\obeta}\right)\right),\\
  \delta \obeta =&-\frac{W_{\obeta}(\delta \phi,\delta \psi;\beta,\obeta)}{W_{\obeta\obeta}(\phi,\psi;\beta,\obeta)} =-\frac{1}{W_{\obeta\obeta}} \left(\frac{1}{2}\int_{\Gamma_1} d\lambda\ \delta \phi(\lambda) (\lambda-\beta)^{-1/2}(\lambda-\obeta)^{-3/2}\right.\\&+\left.
\int_{\Gamma_2} d\lambda\ \delta \psi(\lambda) (\lambda-\beta)^{-1/2}(\lambda-\obeta)^{-3/2} 
\left(\frac{1}{2}\ln\left(\frac{\beta-\obeta}{(\lambda-\beta)(\lambda-\obeta)}\right) -\frac{\lambda-\beta}{\beta-\obeta}\right)\right)
 \end{split}
\end{equation}
where 
\begin{equation}
\begin{split}
 W_{\beta\beta}=& \frac{3}{4} \int_{\Gamma_1} d\lambda\ \phi(\lambda) (\lambda-\beta)^{-5/2}(\lambda-\obeta)^{-1/2}\\&+
\int_{\Gamma_2} d\lambda\ \psi(\lambda) (\lambda-\beta)^{-5/2}(\lambda-\obeta)^{-1/2} 
\left(\frac{3}{4}\ln\left( \frac{(\lambda-\beta)(\lambda-\obeta)}{\beta-\obeta}\right) +\frac{(\lambda-\obeta)(6\beta-4\obeta-2\lambda)}{2(\beta-\obeta)^2}\right),\\
 W_{\obeta\obeta}=& \frac{3}{4} \int_{\Gamma_1} d\lambda\ \phi(\lambda) (\lambda-\obeta)^{-5/2}(\lambda-\beta)^{-1/2}\\&+
\int_{\Gamma_2} d\lambda\ \psi(\lambda) (\lambda-\obeta)^{-5/2}(\lambda-\beta)^{-1/2} 
\left(\frac{3}{4}\ln\left( \frac{(\lambda-\beta)(\lambda-\obeta)}{\beta-\obeta}\right) +\frac{(\lambda-\beta)(6\obeta-4\beta-2\lambda)}{2(\beta-\obeta)^2}\right).
\end{split}
\end{equation}
The expression (\ref{deltabeta}) can be written equivalently as
\begin{equation}
 \begin{split}
  \frac{\delta \beta}{\delta \phi(\lambda)} =&-\frac{1}{2W_{\beta\beta}}
(\lambda-\beta)^{-3/2}(\lambda-\obeta)^{-1/2},\\
  \frac{\delta \beta}{\delta \psi(\lambda)} =& -\frac{1}{W_{\beta\beta}} (\lambda-\beta)^{-3/2}(\lambda-\obeta)^{-1/2} 
\left(\frac{1}{2}\ln\left( \frac{\beta-\obeta}{(\lambda-\beta)(\lambda-\obeta)}\right) +\frac{\lambda-\obeta}{\beta-\obeta}\right),\\
  \frac{\delta \obeta}{\delta \phi(\lambda)} =& -\frac{1}{2W_{\obeta\obeta}}
(\lambda-\beta)^{-1/2}(\lambda-\obeta)^{-3/2},\\
  \frac{\delta \obeta}{\delta \psi(\lambda)} =& -\frac{1}{W_{\obeta\obeta}} (\lambda-\beta)^{-1/2}(\lambda-\obeta)^{-3/2} 
\left(\frac{1}{2}\ln\left(\frac{\beta-\obeta}{(\lambda-\beta)(\lambda-\obeta)}\right) -\frac{\lambda-\beta}{\beta-\obeta}\right).
 \end{split}
\end{equation}
These formulae describe the variation of positions of critical points $(\beta,\obeta)$ for arbitrary variations $\delta \phi(\lambda)$ and $\delta \psi(\lambda)$ of functions $\phi(\lambda)$ and $\psi (\lambda)$ with the fixed contours $\Gamma_1$ and $\Gamma_2$. Choosing specific classes of functions  $\phi(\lambda)$ and $\psi (\lambda)$  one gets explicit formulae for variations of $\beta$ and $\obeta$.\par
As a consequence of these relations one has
\begin{prop}
 For generic critical points and arbitrary infinitesimal variations $\delta \phi(\lambda)$, $\delta \psi(\lambda)$ of
functions $\phi(\lambda)$ and $\psi(\lambda)$
\begin{equation}
\label{WpotentialF}
 W(\delta \phi(\lambda),\delta \psi(\lambda); \beta,\obeta)=\delta F(\beta,\obeta).
\end{equation}
\end{prop}
{\bf Proof} Formulae (\ref{deltabeta}) imply that for an arbitrary infinitesimal variations $\delta \phi$, $\delta \psi$, 
\begin{equation}
 \frac{\delta \beta}{ W_\beta(\delta \phi,\delta \psi)}=-\frac{1}{W_{\beta\beta}(\phi,\psi)}, \qquad
 \frac{\delta \obeta}{ W_\obeta(\delta \phi,\delta \psi)}=-\frac{1}{W_{\obeta\obeta}(\phi,\psi)}.
\end{equation}
For another variation $\delta' \phi$, $\delta' \psi$ one has
\begin{equation}
 \frac{\delta' \beta}{ W_\beta(\delta' \phi,\delta' \psi; \beta,\obeta)}=-\frac{1}{W_{\beta\beta}(\phi,\psi; \beta,\obeta)}, \qquad
 \frac{\delta' \obeta}{ W_\obeta(\delta' \phi,\delta' \psi; \beta,\obeta)}=-\frac{1}{W_{\obeta\obeta}(\phi,\psi; \beta,\obeta)}.
\end{equation}
So, for two arbitrary variations $\delta \phi$, $\delta \psi$ and $\delta' \phi$, $\delta' \psi$ the following
relations are valid
\begin{equation}
\label{W-1form-tech}
\begin{split}
  W_\beta(\delta' \phi,\delta' \psi; \beta,\obeta) \delta \beta =&
W_\beta(\delta \phi,\delta \psi; \beta,\obeta) \delta' \beta,  \\
  W_\obeta(\delta' \phi,\delta' \psi; \beta,\obeta) \delta \obeta =&
W_\obeta(\delta \phi,\delta \psi; \beta,\obeta) \delta' \obeta.  
\end{split}
\end{equation}
Summing up the relations (\ref{W-1form-tech}) one gets
\begin{equation}
\delta  W(\delta' \phi,\delta' \psi; \beta,\obeta) =
\delta' W(\delta \phi,\delta \psi; \beta,\obeta).
\end{equation}
Thus locally there exists a function $F(\beta,\obeta)$ such that (\ref{WpotentialF}) holds. $\square$\par
Representing infinitesimal variations $\delta \phi(\lambda)$ and $\delta \psi(\lambda)$ as
\begin{equation}
\label{phidiscrvar}
 \delta \phi(\lambda)=\sum_{k=1}^n dx_k \phi_k(\lambda), \qquad 
\delta \psi(\lambda)=\sum_{k=1}^m dy_k \psi_k(\lambda)
\end{equation}
where $x_k$, $y_k$ are parameters, $\phi_k(\lambda)$, $\psi_k(\lambda)$ are certain functions and $n$, $m$ are certain integers, one can rewrite relations (\ref{WpotentialF})  as
\begin{equation}
\label{W-exact}
 \sum_{k=1}^n  W(\phi_k,0;\beta,\obeta)dx_k+\sum_{l=1}^m  W(0,\psi_k;\beta,\obeta)dy_l=dF(\beta,\obeta).
\end{equation}
So
\begin{equation}
\label{WFx}
  W(\phi_k,0;\beta,\obeta)=F_{x_k}, \qquad  W(0,\psi_l;\beta,\obeta)=F_{y_l}, \qquad k=0,\dots,n,\quad l=0,\dots,m.
\end{equation}
We emphasize that in these formulae the functions $\phi$ and $\psi$ are not necessarily linear in variables $x_k$, $y_k$.\par
Let us consider the equations (\ref{deltabeta}) defining variation of $\beta$ and $\obeta$. For the infinitesimal variations given by (\ref{phidiscrvar})the formulae (\ref{deltabeta}) imply the differential equations
\begin{equation}
\label{bxby}
 \begin{split}
&\beta_{x_k}=-\frac{W_\beta(\phi_k,0;\beta,\obeta)}{W_{\beta \beta}(\phi,\psi;\beta,\obeta)}, \qquad 
 \obeta_{x_k}=-\frac{W_\obeta(\phi_k,0;\beta,\obeta)}{W_{\obeta \obeta}(\phi,\psi;\beta,\obeta)}, \quad k=0,\dots,n\\
&\beta_{y_l}=-\frac{W_\beta(0,\psi_l;\beta,\obeta)}{W_{\beta \beta}(\phi,\psi;\beta,\obeta)}, \qquad 
 \obeta_{y_l}=-\frac{W_\obeta(0,\psi_l;\beta,\obeta)}{W_{\obeta \obeta}(\phi,\psi;\beta,\obeta)}, \quad l=0,\dots,m.
 \end{split}
\end{equation}
These equations define the dependence of the critical points $\beta,\,\obeta$  on each deformation parameter $\phi_k$ (or $\psi_l$) $via$ an ordinary differential equation. For $n,m>1$ the r.h.s. of equations (\ref{bxby}) depend on the other deformation parameters through the functions $\beta(x,y)$, $\obeta(x,y)$. For the functions $W(\phi,\psi;\beta,\obeta)$ with $\phi,\psi$ of the form (\ref{philin}) these r.h.s. depend on all $x$ and $y$ explicitly. So, for $n,m>1$ it is quite natural to look for the common solution of equations (\ref{bxby}). Such solutions obey the system of quasilinear PDEs of the form
\begin{equation}
\label{bxbyPDE}
 \begin{split}
 & \frac{\beta_{x_k}}{W_\beta(\phi_k,0;\beta,\obeta)}=\frac{\beta_{x_l}}{W_\beta(\phi_l,0;\beta,\obeta)}, \qquad
 \frac{\obeta_{x_k}}{W_\obeta(\phi_k,0;\beta,\obeta)}=\frac{\obeta_{x_l}}{W_\obeta(\phi_l,0;\beta,\obeta)}, \quad
k,l=1,\dots,n \\ 
& \frac{\beta_{y_k}}{W_\beta(0,\psi_k;\beta,\obeta)}=\frac{\beta_{y_l}}{W_\beta(0,\psi_k;\beta,\obeta)}, \qquad
 \frac{\obeta_{y_k}}{W_\obeta(0,\psi_k;\beta,\obeta)}=\frac{\obeta_{y_l}}{W_\obeta(0,\psi_k;\beta,\obeta)}, \quad
k,l=1,\dots,m 
 \end{split}
\end{equation}
and ``mixed'' equation
 \begin{equation}
\label{bxbyPDE-mix}
 \begin{split}
 & \frac{\beta_{x_k}}{W_\beta(\phi_k,0;\beta,\obeta)}=\frac{\beta_{y_l}}{W_\beta(0,\psi_l;\beta,\obeta)}, \qquad
 \frac{\obeta_{x_k}}{W_\obeta(\phi_k,0;\beta,\obeta)}=\frac{\obeta_{y_l}}{W_\obeta(0,\phi_l;\beta,\obeta)}, \quad
k=1,\dots,n,\ l=1,\dots,m\ .
 \end{split}
\end{equation}
It is a simple direct check that for the common solutions $\beta(x,y)$, $\obeta(x,y)$ of these equations 
due to the relation (\ref{bxby}) one has  the conservation laws
\begin{equation}
 \label{claw}
\frac{\partial W(\phi_k,0;\beta,\obeta)}{\partial x_l}=\frac{\partial W(\phi_l,0;\beta,\obeta)}{\partial x_k}, \qquad
\frac{\partial W(0,\psi_k;\beta,\obeta)}{\partial x_l}=\frac{\partial W(0,\psi_l;\beta,\obeta)}{\partial x_k}. 
\end{equation}
Equations (\ref{bxbyPDE}) represent themselves  the hydrodynamic type systems with $\beta$ and $\obeta$ being the Riemann invariants and the characteristic velocities 
\begin{equation}
 \lambda_{k,l}(\beta,\obeta)= \frac{W_\beta(\phi_k,0;\beta,\obeta)}{W_\beta(\phi_l,0;\beta,\obeta)}, \qquad
 \overline{\lambda}_{k,l}(\beta,\obeta)= \frac{W_\obeta(0,\psi_k;\beta,\obeta)}{W_\obeta(0,\phi_l;\beta,\obeta)}.
\end{equation}
All equations (\ref{bxby}), (\ref{bxbyPDE}), (\ref{bxbyPDE-mix}) are compatible by construction. Since the space of arbitrary functions $\phi(\lambda)$ and
$\psi(\lambda)$  is infinite-dimensional, one has in general infinite families (hierarchies) of equations (\ref{bxby})  and (\ref{bxbyPDE}). The function $W$ is the generating function for all these equations.\par
For the particular class of functions $\phi$ and $\psi$ representable as a linear superposition
\begin{equation}
\label{philin}
   \phi(\lambda)=\sum_{k=1}^N x_k \phi_k(\lambda), \qquad 
 \psi(\lambda)=\sum_{k=1}^M  y_k \psi_k(\lambda)
\end{equation}
where $\phi_k(\lambda)$ and $\psi_k(\lambda)$ form a basis for functions of this class, 
the function $W$ is of the form
\begin{equation}
\label{W-linxy}
 W(x,y;\beta,\obeta)=\sum_{k=1}^N x_k W_k(\beta,\obeta)+\sum_{k=1}^M y_k \widetilde{W_k}(\beta,\obeta)
\end{equation}
where 
\begin{equation}
 \begin{split}
  W_k(\beta,\obeta)=&\int_{\Gamma_1} d\lambda\ \phi_k(\lambda) (\lambda-\beta)^{-1/2}(\lambda-\obeta)^{-1/2}, \\
\widetilde{W_l}(\beta,\obeta)=&\int_{\Gamma_2} d\lambda\ \psi_l(\lambda) (\lambda-\beta)^{-1/2}(\lambda-\obeta)^{-1/2} \ln\left( \frac{\beta-\obeta}{(\lambda-\beta)(\lambda-\obeta)}\right),
 \end{split}
\end{equation}
and the formula (\ref{W-exact}) becomes
\begin{equation}
\label{deWdF}
  \delta W(x,y;\beta,\obeta)=W(x+dx,y+dy;\beta,\obeta)-W(x,y;\beta,\obeta)=W(dx,dy;\beta,\obeta)=dF(\beta,\obeta)
\end{equation}
and one has analogs of the formulae (\ref{WFx}).
In fact for the function $W$ of the form (\ref{W-linxy})  it is a simple consequence of general properties of its differential. Indeed, the differential 
\begin{equation}
 d W(x,y;z,\oz)
=W_z dz +W_\oz d\oz +\sum_{k} W_{x_k} dx_k+\sum_{k} W_{x_k} dy_k
\end{equation}
evaluated at the critical point $W_\beta=W_\obeta=0$ is
\begin{equation}
  d W(x,y;\beta,\obeta)= \sum_{k} W_{x_k}(\beta,\obeta) dx_k+\sum_{k} W_{x_k}(\beta,\obeta) dy_k =W(dx,dy;\beta,\obeta)=dF(x,y,\beta,\obeta)
\end{equation}
where $F(x,y,\beta,\obeta)=W(x,y,z,\oz)\Big{|}_{z=\beta,\ \oz=\obeta}$.
\par
In a different situation, namely for the versal deformation of the classical ADE singularities, the relation of the type (\ref{deWdF}) has been discussed in \cite{Kon}.\par
We note that for the particular class of functions given by the formula (\ref{philin}), the function $W(\phi,\psi;\beta,\obeta)$ essentially coincides with the generating function $S(t,\beta)$ of dispersionless hierarchies (see e.g. \cite{Kri}).\par
At last there is a natural question on the possible connection between the PDEs describing the critical points of function $W$ and those for dual function $W^*$defined by the relations (\ref{W*}). In fact one has
\begin{prop}
\label{WW*-prop}
 Differential equations which describe the dependence on $x_i,y_i$ of the critical points $\beta,\obeta$ for the functions $W(x,y;z,\oz)$ of the form (\ref{W-linxy}) coincide with those for the corresponding dual function $W^*(x,y;z,\oz)$.
\end{prop}
{\bf Proof} First, due to the relations (\ref{W*}), the dual function $W^*$ also has the form  (\ref{W-linxy}), i.e.
\begin{equation}
 W^*(x,y;z,\oz) = \sum_{k=1}^N x_k W_k^*(z,\oz)+\sum_{k=1}^M y_k \widetilde{W}_k^*(z,\oz)
\end{equation}
where $W_k^*$ and $\widetilde{W}_k^*$ are certain functions and, also
\begin{equation}
\label{W*klin}
 \begin{split}
  &{W^*_k}_z=(z-{\oz}) {W_k}_z, \qquad  {W^*_k}_{\oz}=(z-\oz) {W_k}_{\oz}, \qquad k=1,\dots,N \\
  &{\widetilde{W}^*_{k\, z}}=(z-\oz) \widetilde{W}_{k\, z}, \qquad  \widetilde{W}^*_{k\, \oz}=(z-{\oz}) \widetilde{W}_{k\, \oz}, \qquad k=1,\dots,M.  
 \end{split}
\end{equation}
At the critical points $(\beta,\obeta)$ one has the relations (\ref{W*klin}) with $z=\beta$ and $\oz=\obeta$. Then, the relations (\ref{W*}) imply that
\begin{equation}
 W^*_{zz}=W_z+(z-\oz)W_{zz},\qquad W^*_{\oz \oz}=W_\oz+(z-\oz)W_{\oz \oz}. 
\end{equation}
So, at the critical points
\begin{equation}
 W^*_{\beta \beta}=(\beta-\obeta)W_{\beta \beta},\qquad W^*_{\obeta \obeta}=(\obeta-\beta)W_{\obeta \obeta}. 
\end{equation}
Consequently 
\begin{equation}
 \frac{W^*_{k\, \beta}}{W^*_{\beta \beta}}=\frac{W_{k\, \beta}}{W_{\beta \beta}}, \qquad
 \frac{W^*_{k\, \obeta}}{W^*_{\obeta \obeta}}=\frac{W_{k\, \obeta}}{W_{\obeta \obeta}},\qquad, k=1,\dots,N
\end{equation}
and
\begin{equation}
 \frac{\widetilde{W}^*_{k\, \beta}}{\widetilde{W}^*_{\beta \beta}}=\frac{\widetilde{W}_{k\, \beta}}{\widetilde{W}_{\beta \beta}}, \qquad
 \frac{\widetilde{W}^*_{k\, \obeta}}{\widetilde{W}^*_{\obeta \obeta}}=\frac{\widetilde{W}_{k\, \obeta}}{\widetilde{W}_{\obeta \obeta}},\qquad, k=1,\dots,M
\end{equation}
In virtue of these relations the r.h.s. of equations (\ref{bxby}), associated with the function $W$ and its dual $W^*$ coincide and, thus the PDEs (\ref{bxbyPDE}), (\ref{bxbyPDE-mix}) coincide as well. $\square$ 
\section{Integrable equations and hierarchies}
Each choice of the functions $\phi(\lambda)$ and $\psi(\lambda)$, their variations and contours $\Gamma_1$ and  $\Gamma_2$ provides us with the concrete examples of equations (\ref{bxbyPDE}) and (\ref{bxbyPDE-mix}). Here, we will discuss some distinguished examples of these equations.\par
First, let us choose the functions $\phi$ and $\psi$ as 
\begin{equation}
 \phi(\lambda)=\sum_{k=1}^\infty x_k \lambda^k, \qquad 
 \psi(\lambda)=\sum_{k=1}^\infty y_k \lambda^k,
\end{equation}
and $\Gamma_1$, $\Gamma_2$ being the circles of large radius. In this case
\begin{equation}
\label{W-lin}
 W(x,y;z,\oz)=\sum_{k=1}^\infty x_k W_k(z,\oz)+\sum_{k=1}^\infty y_k \widetilde{W}_k(z,\oz)
\end{equation}
where
\begin{equation}
\begin{split}
 W_k(z,\oz)=&2\pi i \mathrm{Res}_{\lambda=\infty}\left( \frac{\lambda^k}{(\lambda-z)^{1/2}(\lambda-\oz)^{1/2}}\right), 
\\
\widetilde{W}_k(z,\oz)=&2\pi i \mathrm{Res}_{\lambda=\infty}\left(\frac{\lambda^k}{(\lambda-z)^{1/2}(\lambda-\oz)^{1/2}}
\ln \left(  \frac{z-\oz}{(\lambda-z)(\lambda-\oz)}  \right)\right), \qquad k=0,1,2,\dots\ .
\end{split}
\end{equation}
So 
 \begin{equation}
 \label{W-std}
 \begin{split}
  \frac{1}{2\pi i} W=& \frac{x_1}{2}(z+\oz)+\frac{x_2}{8}(3z^2+2z\oz+3\oz^2)+\frac{x_3}{16}(5z^3+3z^2\oz+3z\oz^2+5\oz^3)+\dots \\
  &+y_0\ln(z-\oz)+\frac{y_1}{2}(2(z+\oz)+(z+\oz)\ln(z-\oz))+\dots\ . 
 \end{split}
 \end{equation}
Critical points $\beta$ of the function (\ref{W-std}) are defined by the equations
 \begin{equation}
\label{Wb-std}
 \begin{split} 
 W_\beta=& \frac{x_1}{2}+\frac{x_2}{4}(3\beta+\obeta)+\frac{3 x_3}{16}(5\beta^2+2 \beta \obeta + \obeta^2)+\dots
  +\frac{y_0}{z-\oz}+y_1 \frac{3\beta-\obeta+(\beta-\obeta)\ln(\beta-\obeta)}{2(\beta-\obeta)}+\dots\ =0,\\
  W_\obeta=& \frac{x_1}{2}+\frac{x_2}{4}(3\obeta+\beta)+\frac{3 x_3}{16}(5\obeta^2+2 \beta \obeta + \beta^2)+\dots
  -\frac{y_0}{z-\oz}+y_1 \frac{\beta-3\obeta+(\beta-\obeta)\ln(\beta-\obeta)}{2(\beta-\obeta)}+\dots\ =0.
\end{split}
 \end{equation}
First family of equations (\ref{bxbyPDE}) with $l=1$ is of the form 
\begin{equation}
\label{dNLS-h}
 \beta_{x_k}=\lambda_k(\beta,\obeta) \beta_{x_1}, \quad \obeta_{x_k}=\overline{\lambda_k}(\beta,\obeta) \obeta_{x_1}, \qquad k=1,2,3,\dots 
\end{equation}
where $\lambda_1=1$, $\lambda_2=\frac{1}{4}(3\beta+\obeta)$, $\lambda_3=\frac{3}{16}(5\beta^2+2 \beta \obeta + \obeta^2)$, \dots\ .
It is the well known dispersionless focusing nonlinear Schr\"{o}dinger (dNLS) hierarchy \cite{Zak} or quasi-classical Da Rios hierarchy \cite{KO1,KO2}.\par
The second family of equations (\ref{bxbyPDE}) with $l=0$ is given by the equations
\begin{equation}
\label{dToda-h}
 \beta_{y_k}=\mu_k(\beta,\obeta) \beta_{y_1}, \quad \obeta_{y_k}=\overline{\mu_k}(\beta,\obeta) \obeta_{y_1}, \qquad k=0,1,2\dots
\end{equation} 
where $\mu_0=1$,  $\mu_1= \frac{(3\beta-\obeta+(\beta-\obeta)\ln(\beta-\obeta)}{2}$, $\overline{\mu_1}=\frac{\beta-3\obeta+(\beta-\obeta)\ln(\beta-\obeta)}{2}$  and so on. Equations (\ref{dToda-h}) coincide with equations found earlier in a completely different setting in \cite{EY}, only rewritten in terms of the Riemann invariants $\beta$ and $\obeta$. \par
The simplest mixed system (\ref{bxbyPDE-mix}) is 
\begin{equation}
\label{dTodab}
 \beta_{x_1}=\frac{\beta-\obeta}{2}\beta_{y_0},\qquad \obeta_{x_1}=-\frac{\beta-\obeta}{2}\obeta_{y_0}
\end{equation}
which is the well known dispersionless Toda (dToda) equation
\begin{equation}
\label{dToda}
 \phi_{x_1x_1}+e^{\phi}_{y_0y_0}=0
\end{equation}
where $\phi=2\ln(i(\beta-\obeta))$ is real. The hierarchy of equations (\ref{dToda-h}) together with the standard d-Toda hierarchy (see \cite{TT}) has been referred in \cite{EY}  as the extended dToda hierarchy. It is natural to refer to the full hierarchy of equations (\ref{bxbyPDE}), (\ref{bxbyPDE-mix}) associated with the function $W$ (\ref{W-lin}) as the extended dNLS/dToda hierarchy. This hierarchy describes the family of critical points of a special class of functions $W$ defined by the formula   (\ref{W-lin}).\par
As far as the equivalence mentioned in the proposition \ref{WW*-prop} is concerned, the hierarchy of quasi-linear equations associated with the function $W$ given by (\ref{W-lin}) with $y_k=0$, $k\geq 1$. 
 (i.e. dToda equation plus dNLS hierarchy) coincide with  dToda hierarchy describing critical points of the function
 \begin{equation}
  W^*=y_0(z+\oz)+\oint_{C_\infty}\frac{d\lambda}{2\pi i} (\lambda-z)^{1/2}(\lambda-\oz)^{1/2} \sum_{n\geq 1}\lambda^{n-1}x_n, 
 \end{equation}
which obeys the EPD equation $E(1/2,1/2)$.\par
This observation clarifies also the appearance of the dToda equation in two apparently different situations one of which is connected 
with the EPD equation $E(-1/2,-1/2)$ \cite{DGK,KMAM2,KMAM3} and the EPD equation $E(1/2,1/2)$  with the function $W$ given by the formula (\ref{W-std}) (see also \cite{KO1,KO2}). \par
Quantities $\widetilde{W}$ in (\ref{W-lin}) are connected with the solutions $h^{(1)}_{k+2}$ of the EPD equation $E(-1/2,-1/2)$ found in \cite{PP}. Namely up to the trivial constants 
\begin{equation}
\label{hcons}
 \frac{\partial^2 h^{(1)}_{k+2}}{\partial \beta \partial \obeta}, \qquad k=0,1,\dots
\end{equation}
where $u=-\beta-\obeta$, $v=\frac{1}{4}(\beta-\obeta)^2$, 
the quantities $h^{(1)}_{k}$ are the conserved densities of the dNLS equation \cite{PP}. So, the formula (\ref{hcons}) establishes the connection between conserved density for equation (\ref{dNLS-h}) and symmetries for equation (\ref{dToda-h}). On the other hand formula
(\ref{hcons}) represents a particular instance of the old known relation 
$\frac{\partial^2 W_\epsilon}{\partial z \partial \oz}=W_{\epsilon+1}$ between the solutions $W_\epsilon$ of the EPD equations $E(\epsilon,\epsilon)$ with different $\epsilon$ \cite{Dar}.\par 
Choosing now the functions  $\phi(\lambda)$ and $\psi(\lambda)$ as
\begin{equation}
  \phi(\lambda)=\sum_{k=1}^\infty x_k \lambda^{-k} , \qquad 
 \psi(\lambda)=\sum_{k=1}^\infty  y_k \lambda^{-k},
\end{equation}
and contours $\Gamma_1$ and  $\Gamma_2$ to be the circles of small radius around the origin, one has 
\begin{equation}
\label{W-lininv}
 W(x,y;z,\oz)=\sum_{k=1}^\infty x_k W_k(z,\oz)+\sum_{k=1}^\infty y_k \widetilde{W}_k(z,\oz)
\end{equation}
where
\begin{equation}
\begin{split}
 W_k(z,\oz)=&-\frac{2\pi i}{(z\oz)^{1/2}}  \mathrm{Res}_{\lambda=0}\left( \frac{\lambda^{-k}}{\left(1-\frac{\lambda}{z}\right)^{1/2}\left(1-\frac{\lambda}{\oz}\right)^{1/2}}\right), 
\\
 \widetilde{W}_k(z,\oz)=&-\frac{2\pi i}{(z\oz)^{1/2}}\ln\left( \frac{z-\oz}{z\oz}\right)W_k(z,\oz)
 \\&+ \frac{2\pi i}{(z\oz)^{1/2}} \mathrm{Res}_{\lambda=0}
 \left( 
 \frac{\lambda^{-k}}{\left(1-\frac{\lambda}{z}\right)^{1/2}\left(1-\frac{\lambda}{\oz}\right)^{1/2}}
\ln \left( \left(1-\frac{\lambda}{z}\right)
\left(1-\frac{\lambda}{\oz}\right) \right)
 \right), \qquad k=0,1,2,\dots\ .
\end{split}
\end{equation}
So 
 \begin{equation}
 \label{W-at-zero}
 \begin{split}
  \frac{1}{2\pi i} W=& -\frac{x_1}{(z\oz)^{1/2}} - \frac{x_2}{2(z\oz)^{1/2}}(\frac{1}{z}+\frac{1}{\oz})
  -\frac{x_2}{8(z\oz)^{1/2}}(\frac{3}{z^2}+\frac{2}{z\oz}+\frac{3}{\oz^2})+\dots \\
  &-\frac{y_1}{(z\oz)^{1/2}}\ln\left(\frac{z-\oz}{z\oz}\right)
  -\frac{y_2}{2(z\oz)^{1/2}}\left(\frac{2}{z}+\frac{2}{\oz}+\left(\frac{1}{z}+\frac{1}{\oz}\right)
   \ln\left(\frac{z-\oz}{z\oz}\right)\right)+\dots\ . 
 \end{split}
 \end{equation}
Differential equations (\ref{bxbyPDE}) for critical points at $l=1$ are of the form (with $z = \beta$)
\begin{equation}
 \begin{split}
   \beta_{x_k}=&\lambda_k(\beta,\obeta) \beta_{x_1}, \quad \obeta_{x_k}=\overline{\lambda_k}(\beta,\obeta) \obeta_{x_1}, \qquad k=0,1,2\dots\\
   \beta_{y_k}=&\mu_k(\beta,\obeta) \beta_{y_1}, \quad \obeta_{y_k}=\overline{\mu_k}(\beta,\obeta) \obeta_{y_1}, \qquad k=0,1,2\dots
 \end{split}
\end{equation}
where $\lambda_1=\overline{\lambda_1}=\mu_1=\overline{\mu_1}=1$ and
\begin{equation}
\begin{split}
 \lambda_2=\frac{3}{2\beta}+\frac{1}{2\obeta}, \qquad \overline{\lambda}_2=\frac{3}{2\obeta}+\frac{1}{2\beta} \\
\mu_2=\frac{2 \left(\beta^2+\beta \obeta-4 \obeta^2\right)+\left(\beta^2+2 \beta
   \obeta-3 \obeta^2\right) \ln
   \left(\frac{1}{\obeta}-\frac{1}{\beta}\right)}{2 \beta \obeta
   \left((\beta-\obeta) \ln
   \left(\frac{1}{\obeta}-\frac{1}{\beta}\right)-2 \obeta\right)},  \\
\overline{\mu}_2= \frac{2 \left(-4 \beta^2+\beta \obeta+\obeta^2\right)+\left(-3 \beta^2+2 \beta
   \obeta+\obeta^2\right) \ln
   \left(\frac{1}{\obeta}-\frac{1}{\beta}\right)}{2 \beta \obeta
   \left((\obeta-\beta) \ln
   \left(\frac{1}{\obeta}-\frac{1}{\beta}\right)-2 \beta\right)}.
\end{split}
\end{equation}
The first mixed equations (\ref{bxbyPDE-mix}) are
\begin{equation}
\frac{-2 \beta_{x_1}}{\beta^{-3/2} \obeta^{-1/2}}= \frac{\beta_{y_1}}{\frac{\obeta \left((\obeta-\beta) \log
   \left(\frac{1}{\obeta}-\frac{1}{\beta}\right)+2
   \obeta\right)}{2 (\beta-\obeta) (\beta \obeta)^{3/2}}} , \qquad 
\frac{-2 \obeta_{x_1}}{\obeta^{-3/2} \beta^{-1/2}}= \frac{\obeta_{y_1}}{\frac{\beta \left((\obeta-\beta) \log
   \left(\frac{1}{\obeta}-\frac{1}{\beta}\right)-2 \beta\right)}{2
   (\beta-\obeta) (\beta \obeta)^{3/2}}}.
\end{equation}
All these equations have simpler form in terms of the new dependent variable $\frac{1}{\beta},\frac{1}{\obeta}$.\par
In fact it is easy to see that the function $W$ in (\ref{W-at-zero}) is connected with the function (\ref{W-std}) by the relation (\ref{Appell}) for the transformation $z \to z'=1/z$ ($a=d=0, b=c=1$), in accordance also with the formulae (\ref{psipsi}) and (\ref{phiphi}). We note that the critical points of the functions $W$ in (\ref{W-at-zero})
and  $W$ in (\ref{W-std}) do not coincide and they are not connected by the transformation $\beta \to -1/\beta$
and $\obeta \to -1/\obeta$.\par
Different class of quasi-linear PDEs is associated with the choice 
\begin{equation}
\label{phidelta}
\phi(\lambda)= \sum_{k=0}^N x_k \delta(\lambda-\lambda_k), \qquad 
\psi(\lambda)= \sum_{k=0}^M y_k \delta(\lambda-\mu_k)
\end{equation}
where $\delta$ is the Dirac delta function, $\lambda_k,\mu_k$ are real coefficients and $\Gamma_1$ and $\Gamma_2$ are
real axis. The corresponding function $W$ is
\begin{equation}
\label{W-delta}
 W(x,y;z,\oz)=\sum_{k=1}^N x_k W_k(z,\oz)+\sum_{k=1}^M y_k \widetilde{W}_k(z,\oz)
\end{equation}
where
\begin{equation}
\begin{split}
 W_k(z,\oz)=&\left(\lambda_k-z\right)^{-1/2} \left(\lambda_k-\oz\right)^{-1/2}, 
\\
 \widetilde{W}_k(z,\oz)=&.\left({\lambda_k}-{z}\right)^{-1/2} \left(\lambda_k-\oz\right)^{-1/2} 
\ln \left( \frac{z-\oz}{(\lambda_k-z)(\lambda_k-\oz)}\right).
\end{split}
\end{equation}
In this case equations (\ref{bxbyPDE}) are of the form
\begin{equation}
\label{bxby-delta}
\begin{split}
& (\lambda_k-\beta)^{3/2}(\lambda_k-\obeta)^{1/2} \beta_{x_k}
=(\lambda_l-\beta)^{3/2}(\lambda_l-\obeta)^{1/2} \beta_{x_l}, \\
& (\lambda_k-\beta)^{1/2}(\lambda_k-\obeta)^{3/2} \beta_{x_k}
=(\lambda_l-\beta)^{1/2}(\lambda_l-\obeta)^{3/2} \beta_{x_l}, \qquad k=1,\dots,N\ l=1,\dots,M.
\end{split}
\end{equation}
Without loss of generality one can choose $\lambda_0=0$. The subsystem of equations (\ref{bxby-delta}) with $l=0$ becomes
\begin{equation}
\label{bxby-delta-0}
\begin{split}
& \beta_{x_k}= \left(1-\frac{\lambda_k}{\beta}\right)^{-3/2}\left(1-\frac{\lambda_k}{\obeta}\right)^{-1/2}\beta_{x_0},\\ 
& \obeta_{x_k}= \left(1-\frac{\lambda_k}{\beta}\right)^{-1/2}\left(1-\frac{\lambda_k}{\obeta}\right)^{-3/2}\obeta_{x_0}, \qquad k=1,2,\dots,N.
\end{split}
\end{equation}
After rescaling $\beta \to \lambda_k \beta $ all these equations assume the form
\begin{equation}
\begin{split}
 & \beta_{x_k}= \left(1-\frac{1}{\beta}\right)^{-3/2}\left(1-\frac{1}{\obeta}\right)^{-1/2}\beta_{x_0},\\ 
& \obeta_{x_k}= \left(1-\frac{1}{\beta}\right)^{-1/2}\left(1-\frac{1}{\obeta}\right)^{-3/2}\obeta_{x_0},
\end{split}
\end{equation}
common for the whole family. In terms of new variable $u=1-\frac{1}{\beta}$ one has the system ($x_k=t$)
\begin{equation}
\label{uuuu}
 u_t=u^{-3/2}\overline{u}^{-1/2} u_{x_0}, \qquad  \overline{u}_t=u^{-1/2}\overline{u}^{-3/2} \overline{u}_{x_0},
\end{equation}
or, equivalently
\begin{equation}
 u_t=\frac{\partial \omega}{\partial u} u_{x_0}, \qquad  
\overline{u}_t=\frac{\partial \omega}{\partial \overline{u}} \overline{u}_{x_0},
\end{equation}
where $\omega=-2(u\overline{u})^{-1/2}$. This system describes critical points of the function 
\begin{equation}
 W=x_0(z+\oz)- 2 t (z\oz)^{-1/2}.
\end{equation}
For the PDEs in the variables $y_k$ one has equations (\ref{bxbyPDE})  with
\begin{equation}
 \begin{split}
  & W_\beta(0,\psi_k;\beta,\obeta)=(\mu_k-\beta)^{-3/2}(\mu_k-\obeta)^{-1/2} \left( \frac{1}{2} \ln \left( \frac{\beta-\obeta}{(\mu_k-\beta)(\mu_k-\obeta)}\right) + \frac{\mu_k-\obeta}{\beta-\obeta}\right) \\
  & W_\obeta(0,\psi_k;\beta,\obeta)=(\mu_k-\beta)^{-1/2}(\mu_k-\obeta)^{-3/2} \left( \frac{1}{2} \ln \left( \frac{\beta-\obeta}{(\mu_k-\beta)(\mu_k-\obeta)}\right) - \frac{\mu_k-\beta}{\beta-\obeta}\right) 
 \end{split}
\end{equation}
while the simplest ``mixed'' equations (\ref{bxbyPDE-mix}) are
\begin{equation}
\label{deltamix}
\begin{split}
& (\lambda_k-\beta)^{-3/2}(\lambda_k-\obeta)^{-1/2} \beta_{x_k}= 
\frac{(\mu_l-\beta)^{3/2}(\mu_l-\obeta)^{1/2}}{ \left(  \ln \left( \frac{\beta-\obeta}{(\mu_l-\beta)(\mu_l-\obeta)}\right) + 2 \frac{\mu_l-\obeta}{\beta-\obeta}\right)  }\beta_{y_k} \\
& (\lambda_k-\beta)^{-1/2}(\lambda_k-\obeta)^{-3/2} \obeta_{x_k}= 
\frac{(\mu_l-\beta)^{1/2}(\mu_l-\obeta)^{3/2}}{ \left(  \ln \left( \frac{\beta-\obeta}{(\mu_l-\beta)(\mu_l-\obeta)}\right) - 2 \frac{\mu_l-\beta}{\beta-\obeta}\right)  }\obeta_{y_k},
\quad k=1,\dots,N,\ l=1,\dots,M.  
\end{split}
\end{equation}
Choosing $\mu_0=0$ and introducing the variable $\gamma=1/\beta$, one gets from equations (\ref{deltamix}) the system
\begin{equation}
\label{deltamix-beta}
\begin{split}
 (1-\lambda_k \gamma)^{3/2}(1-\lambda_k \overline{\gamma})^{1/2} \gamma_{x_k}=&  
 \frac{\gamma-\overline{\gamma}}{(\gamma-\overline{\gamma})\ln(\gamma-\overline{\gamma})+2\gamma}\gamma_{y_0}\\
 (1-\lambda_k \gamma)^{1/2}(1-\lambda_k \overline{\gamma})^{3/2}  \overline{\gamma}_{x_k}=&  
 \frac{\gamma-\overline{\gamma}}{(\gamma-\overline{\gamma})\ln(\gamma-\overline{\gamma})-2 \overline{\gamma}} \overline{\gamma}_{y_0}
 , \qquad k=1,2,\dots,N.   
\end{split}
 \end{equation}
At $N,M=\infty$ one ha infinite hierarchy of equations (\ref{bxby-delta}), (\ref{bxby-delta-0}), (\ref{uuuu}), (\ref{deltamix}), (\ref{deltamix-beta}). \par
For infinite $N$ and $M$ the is a particularly interesting choice of function $\phi$ and $\psi$ of the type (\ref{phidelta}). It is given by 
\begin{equation}
 \begin{split}
  \phi(\lambda)=-a \sum_{(n,m)\neq(0,0)}\delta(n \lambda+ m), \qquad \psi(\lambda)=-b \sum_{(n,m)\neq(0,0)}\delta(n \lambda+ m),
 \end{split}
\end{equation}
where the sums runs over all pairs of integers $(n,m)\neq(0,0)$ and $a$ and $b$ are arbitrary real constants. So,
\begin{equation}
\label{W-Eis}
 W=a \sum_{(n,m)\neq(0,0)} (nz+m)^{-1/2}(n\oz+m)^{-1/2}+
 b  \sum_{(n,m)\neq(0,0)} (nz+m)^{-1/2}(n\oz+m)^{-1/2} \ln \left( \frac{n^2(z-\oz)}{(nz+m)(n\oz+m)}\right).
\end{equation}
It is a simple check that first term in (\ref{W-Eis}) obeys the relation (\ref{aut12}) with integer $a,b,c,d$. So it is an automorphic form of weight $1/2$. Then under certain additional constraints ( behavior at the cusps of the fundamental domain \cite{Maa1,Maa2})  the function 
\begin{equation}
 W= \sum_{(n,m)\neq(0,0)} (nz+m)^{-1/2}(n\oz+m)^{-1/2}
\end{equation}
is the Maass real analytic modular form $E(z,\oz;1/2)$ or the real analytic Eisenstein series of weight $1/2$ \cite{Maa1,Maa2}. The second term in (\ref{W-Eis}) does not obey (\ref{aut12}) due to the transformation law (\ref{AT}).\par
Now let us consider the function $W$ of the form 
\begin{equation}
 W(x,y;z,\oz)=x(z+\oz)+y E(z,\oz;1/2)
\end{equation}
where $x$ and $y$ are real parameters. Critical points of this function are defined by the equation
\begin{equation}
 W_\beta=x+y \frac{\partial E(\beta,\obeta;1/2)}{\partial \beta}=0, \qquad 
 W_\obeta=x+y \frac{\partial E(\beta,\obeta;1/2)}{\partial \obeta}=0.
\end{equation}
The corresponding PDEs for $\beta$ and $\obeta$ are of the form
\begin{equation}
 \beta_y=\frac{\partial E(\beta,\obeta;1/2)}{\partial \beta} \beta_x, \qquad
  \obeta_y=\frac{\partial E(\beta,\obeta;1/2)}{\partial \obeta} \obeta_x.
\end{equation}
Choosing the function $W$ as the sum
\begin{equation}
 W=W_s+\tilde{y}E(z,\oz;1/2)
\end{equation}
where $W_s$ is given by (\ref{W-std}) one gets an infinite hierarchy of hydrodynamic type systems for which the characteristic velocities are functions of the components of the gradient of the real analytic Eisenstein series $E(z,\oz;1/2)$.\par
Considering instead of $E(z,\oz;1/2)$ any solution of equation (\ref{EPD1212}) obeying (\ref{aut12}), one constructs integrable equations with characteristic velocities containing derivatives of automorphic function of weight $1/2$. This type of equations will be discussed  elsewhere.\par
Within completely different approach the hydrodynamic systems somehow connected with modular forms have been studied in \cite{OS,FO}.
\section{Bi-Hamiltonian structure for EPD E(1/2,1/2) equations hierarchy.}
Peculiarity of the EPD E$(1/2,1/2)$ equations and associated integrable systems is exhibited also in the corresponding Hamiltonian and bi-Hamiltonian structures. The dNLS hierarchy (\ref{dNLS-h}) and the extended dToda hierarchy (\ref{dToda-h}) are known to be bi-Hamiltonian. The corresponding first and second Hamiltonian operators $J_0$ and $J_1$ have the simplest form in the variables
$u$ and $\rho$ defined by $\beta=u+\frac{i}{2}\rho^{1/2}$, namely
\begin{equation}
 J_0=
\left(
\begin{array}{cc}
 0 &  \partial_x \\
 \partial_x  & 0.
\end{array}
\right), \qquad
J_1=
\left(
\begin{array}{cc}
 \rho \partial_x + \partial_x \rho & u \partial_x \\
 \partial_x u & -2\partial_x.
\end{array}
\right), 
\end{equation}
and the first system (\ref{dNLS-h}) is ($t=x_2$, $x=x_1$)
\begin{equation}
 \rho_t=(u\rho)_x,\qquad u_t=-\rho_x+uu_x.
\end{equation}
The pencil $J_1-\lambda J_0$ has a very particular property: the tensors $J_0$ and $J_1$ are resonant since they have a common Casimir $u$ (to recall a Casimir is a function $C$ such that $J \nabla C=0$, where $\nabla C=(\partial_\rho C,\partial_u C)^t$) (see e.g. \cite{Pav}).
As it is known such a property of $J_0$ and $J_1$ poses a problem within the standard Lenard-Magri scheme since this common Casimir cannot be chosen as a starting point for a recursion.\par
Another important fact is that the series of integrals of motion for integrable equations associated with $E(k,k)$ with $k=1/2+\epsilon$, $\epsilon \neq 0$ given in \cite{Nut,NP} collapse in one at the simple naive limit $\epsilon \to 0$.\par
Such a property of integrals of motion and resonant character of the Hamiltonian operators $J_0$ and $J_1$ is clearly a counterpart of the triviality of the naive limit $\epsilon \to 0$ of the general solution of $E(1/2+\epsilon,1/2+\epsilon)$ equation discussed in Section $2$.\par
Poisson's trick again suggests a way to resolve the problem. Let us start with the hierarchy of integrable equations associated
 with $E(1/2+\epsilon,1/2+\epsilon)$. It has compatible Hamiltonian operators
\begin{equation}
 J_0=
\left(
\begin{array}{cc}
 0 &  \partial_x \\
 \partial_x  & 0.
\end{array}
\right), \qquad
 J_1^{\epsilon}=
\left(
\begin{array}{cc}
 \rho \partial_x + \partial_x \rho & u \partial_x +\epsilon \partial_x u  \\
 \epsilon u \partial_x  + \partial_x u & -\rho^{\epsilon}\partial_x- \partial_x \rho^{\epsilon}
\end{array}
\right).
\end{equation}
For the hyperbolic version of the operator see (\cite{Nut}). The important property of $J_0$ and $J_1^\epsilon$ is that at
$\epsilon \neq 0$ they do not suffer of the resonant phenomenon and the Lenard-Magri procedure works for both Casimirs.\par
There are two infinite families of conserved densities $\{H_k,k\geq 0\}$ and $\{\tilde{H}_k,k\geq 0\}$ (see \cite{Nut} and
\cite{NP} for hyperbolic version). The function $\tilde{H}_{1}=u$ is the Casimir density of $J_0$ but not for $J_1^\epsilon$ in general. At $\epsilon=0$
$H_k=\tilde{H}_k$, for every $k \geq 0$ \cite{Nut,NP}.\par
In the naive limit $\lim_{\epsilon \to 0}J_1^{\epsilon} \nabla u=0$ Poisson's observation suggests to take the rescaled density 
$u/\epsilon$ then to consider the flow $X^\epsilon_1=J_1^{\epsilon} \nabla (u/\epsilon)$ and then to calculate the limit $\epsilon \to 0$. One gets
\begin{equation}
\label{ffl}
 X_1 \equiv \lim_{\epsilon \to 0} X^\epsilon_1 =
 \left(
 \begin{array}{c}
 u \\ -\ln\rho
 \end{array}\right)_x
\end{equation}
It is a simple check that this flow can be represented as 
\begin{equation}
 X_1J_0\nabla (H_1^{Toda})
\end{equation}
where $H_1^{Toda} = -\rho(\ln \rho-1)+u^2/2$ is the standard Hamiltonian for the dToda equation.\par
Analogously the straightforward calculations shows that 
\begin{equation}
 \lim_{\epsilon \to 0} X^\epsilon_{n+1} = \lim ((J_1^\epsilon J_0^{-1})^n J_1^\epsilon \nabla (u/\epsilon))=J_0\nabla (H_n^{Toda})+
 J_0\nabla(H_n^{NLS}/\epsilon)
\end{equation}
where $H_n^{Toda}$ are higher nonpolynomial conserved densities for dToda equation and $H_n^{NLS}$ are polynomial densities of the general extended dToda/dNLS hierarchy ($H_n^{dNLS}=H_n$). Thus 
\begin{equation}
\label{relation}
 J_0\nabla (H_n^{Toda})=\lim_{\epsilon \to 0} \left( (J_1^\epsilon J_0^{-1})^n J_1^\epsilon \nabla (u/\epsilon) -
 J_0 \nabla(H_n/\epsilon)\right).
\end{equation}
The first flow (\ref{ffl}) gives the system 
\begin{equation}
\left(
\begin{array}{c}
  \rho \\ u
 \end{array}
\right)_t=
\left(
\begin{array}{c}
  u \\ -\ln \rho
 \end{array}
\right)_x
 \end{equation}
which is, in term of $\phi=\ln \rho$, the dToda equation (\ref{dToda}).  Higher flows $X_n=\lim_{\epsilon \to 0} X_n^\epsilon$ provide us with higher dToda equation.\par The formula (\ref{relation}) clarifies also the specific relation between the Hamiltonian structures for extended dToda/dNLS hierarchy found in \cite{CDZ}.
  \section{Applications to quasiclassical Da Rios system.}
  Study of the vortex filament dynamics in quasi-classical approximation \cite{KO1,KO2} is one of the most important applications of the analysis
  given in previous sections. Here we will indicate only some basic points. \par
  For vortex filament dynamics $\beta=-\tau+iK$ and the quasi-classical Da Rios system (or dispersionless focusing NLS equation) is
  \begin{equation}
\label{bin-intrinsic}
\begin{split}
K_t&+2K_x \tau +K \tau_x=0,\\
\tau_t &- KK_x +2\tau\tau_x=0.
\end{split}
\end{equation}
First higher polynomial flow is
\begin{equation}
\label{KT-sys} 
\begin{split}
K_{t_2} &  -\left( 3\tau^2-\frac{3}{2}K^2\right)  K_x -3 \tau K  \tau_x =0,
 \\
  \tau_{t_2} & +3 \tau K  K_x -\left( 3\tau^2-\frac{3}{2}K^2\right)  \tau_x =0
   \end{split}
\end{equation}
while dToda system is 
\begin{equation}
\label{dTKT}
\begin{split}
 K_{t_{-1}}&-\frac{\tau_x}{K}=0,\\
 K\tau_{t_{-1}}&+\frac{K_x}{K}=0
\end{split}
 \end{equation}
or equation (\ref{dToda}) with $\phi=2\ln(-2K)$ and higher logarithmic system is given by
\begin{equation}
\begin{split}
\label{dT-2KT}
 K_{t_{-2}}&+(2+2\ln(K))K_x+\frac{\tau}{K}\tau_x=0,\\
 \tau_{t_{-2}}&-\frac{\tau}{K}K_x+(2+2\ln(K))\tau_x=0.
\end{split}
 \end{equation}
Two compatible Hamiltonian structures are
\begin{equation} 
J_0= -\left(\begin{array}{ccc} 
0
&&  \partial \frac{1}{K}
\\&&\\   \frac{1}{K} \partial
&& 0
\end{array}\right), \qquad
J_1= - \left(\begin{array}{ccc} 
 \partial  
&& - \partial \frac{\tau}{K}
\\&&\\ -  \frac{\tau}{K} \partial
&&   -\partial
\end{array}\right). 
 \end{equation}
 In the applications to the vortex filament dynamics the function $W$ should be real valued $W(z,\oz)=\overline{W(z,\oz)}$ \cite{KO1}. 
 So a general solution of the quasi-classical Da Rios system (\ref{bin-intrinsic}) is connected with the critical points of the  function
 \begin{equation}
 \label{WDR}
 \begin{split}
  W=&\frac{x}{2}(z+\oz)+\frac{t}{8}(3z^2+2z\oz+3\oz^2)+
  \int_{\mathbb{R}} d\lambda\ \phi(\lambda) (\lambda-z)^{-1/2}(\lambda-\oz)^{-1/2} \\
&+\int_{\mathbb{R}} d\lambda\ \psi(\lambda) (\lambda-z)^{-1/2}(\lambda-\oz)^{-1/2} \ln\left( \frac{z-\oz}{2i(\lambda-z)(\lambda-\oz)}
\right)
 \end{split}
 \end{equation}
where $\phi$ and $\psi$ are two arbitrary real valued functions. \par
 Equation $W_\tau=W_K=0$ for critical points,the hodograph equations for the system  (\ref{bin-intrinsic}) provide us with the family of solutions of the quasi-classical Da Rios system parameterized by two arbitrary real-valued functions of a single argument.
 Evaluating equations $W_\tau=W_K=0$ at $t=0$, one gets
 \begin{equation}
 \begin{split}
  &\int_\mathbb{R} d\lambda\  
\left[\frac{\lambda+\tau_0}{((\lambda+\tau_0)^2+K_0^2)^{3/2}} \left(\phi(\lambda) + \psi(\lambda) \ln\left(\frac{e^{2}K_0}{(\lambda+\tau_0)^2+K_0^2}\right)\right)\right]
=x, \\
& \int_\mathbb{R} d\lambda\  
\left[\frac{K_0}{((\lambda+\tau_0)^2+K_0^2)^{3/2}} \left(\phi(\lambda) + \psi(\lambda) \ln\left(\frac{e^{\frac{K_0^2-(\lambda+\tau_0)^2}{K_0^2}}K_0}{(\lambda+\tau_0)^2+K_0^2}\right)\right)\right]
=0
  \end{split}
 \end{equation}
where $\tau_0=\tau(x,t=0)$ and $K_0=K(x,t=0)$. This system of integral linear equations establishes the relation between the initial data $\tau_0$ and $K_0$ for the Da Rios system and functions $\phi(x)$ and $\psi(x)$. Thus the  functions $\phi(x)$ and $\psi(x)$ 
provide us the parameterization of the infinite family of initial data for the quasi-classical Da Rios system.We would like to emphasize that the complete parameterization of the initial data for the system (\ref{bin-intrinsic}) in terms of the function $W$ require the presence of the second logarithmic term in (\ref{WDR}) with the function $\psi(\lambda)$.\par
For the vortex filament dynamics this means that the higher dNLS (like dispersionless Hirota equation) which corresponds to the non-logarithmic terms in $W$ are not sufficient to describecompletelythe whole family of initial data. Contribution of the dToda flow (\ref{dTKT}), flow (\ref{dT-2KT}) and other flows from the extended dispersionless Toda hierarchy which give large contribution at small curvature \cite{KO2} are essential. In more details this problem will be considered elsewhere.
 \subsubsection*{Acknowledgments}
 The authors thanks G. Carlet, E. Ferapontov and M. Pavlov for useful discussions and informations.  This work was partially supported by the PRIN 2010/11 grant 2010JJ4KBA\_003.

\end{document}